\documentclass[fleqn,usenatbib]{mnras}

\usepackage[T1]{fontenc}

\DeclareRobustCommand{\VAN}[3]{#2}
\let\VANthebibliography\thebibliography
\def\thebibliography{\DeclareRobustCommand{\VAN}[3]{##3}\VANthebibliography}

\usepackage{graphicx}	
\usepackage{amsmath}	
\usepackage{amssymb}
\usepackage{mathtools}
\usepackage{bm}
\usepackage{amsfonts}
\usepackage{subfigure}
\usepackage{slantsc}
\usepackage{soul}

\usepackage[scr=boondox]{mathalfa}
\usepackage{enumitem}

\usepackage{mathtools}

\usepackage{physics}

\usepackage{relsize,exscale}
\usepackage{array}
\usepackage{ctable}
\usepackage{cellspace} 
\newcolumntype{?}{!{\vrule width 0.12em}}

\newcommand{\prospect}{\textsc{prospect}}
\newcommand{\class}{\textsc{class}}
\newcommand{\camb}{\textsc{camb}}
\newcommand{\connect}{\textsc{connect}}
\newcommand{\montepython}{\textsc{MontePython}}
\newcommand{\cobaya}{\textsc{cobaya}}
\newcommand{\getdist}{\textsc{GetDist}}

\newcommand{\CLASS}{\textsc{class}}
\newcommand{\CAMB}{\textsc{camb}}

\renewcommand{\d}{\text{d}}
\newcommand{\thetajfixed}{\tilde{\theta_{j}}}

\newcommand{\appropto}{\mathrel{\vcenter{
			\offinterlineskip\halign{\hfil$##$\cr
				\propto\cr\noalign{\kern2pt}\sim\cr\noalign{\kern-2pt}}}}}

\renewcommand{\vec}[1]{\boldsymbol{#1}}

\title[PROSPECT: A profile likelihood code for cosmology]{PROSPECT: A profile likelihood code for frequentist cosmological parameter inference}

\author[E. B. Holm et al.]{
	Emil Brinch Holm,$^{1}$\thanks{E-mail: emil.brinch@hotmail.com}
	Andreas Nygaard,$^{1}$\thanks{E-mail: andreas@phys.au.dk}
	Jeppe Dakin$^{2}$\thanks{E-mail: jeppe.dakin@uzh.ch}
	Steen Hannestad$^{1}$\thanks{E-mail: sth@phys.au.dk}
	and Thomas Tram$^{1}$\thanks{E-mail: thomas.tram@phys.au.dk}
	\\
	$^{1}$Department of Physics and Astronomy, Aarhus University, DK-8000 Aarhus C, Denmark\\
	$^{2}$Institute for Computation Science, University of Zurich, Winterthurerstrasse 190, 8057 Zürich, Switzerland
}

\date{Accepted 2024 November 12. Received 2024 November 7; in original form 2024 July 24}

\pubyear{2024}

\begin{document}
\label{firstpage}
\pagerange{\pageref{firstpage}--\pageref{lastpage}}
\maketitle

\begin{abstract}
Cosmological parameter inference has been dominated by the Bayesian approach for the past two decades, primarily due to its computational efficiency.
However, the Bayesian approach involves integration of the posterior probability and therefore depends on both the choice of model parametrisation and the choice of prior on the model parameter space. 
In some cases, this can lead to conclusions which are driven by choice of parametrisation and priors rather than by data. The profile likelihood method provides a complementary frequentist tool which can be used to investigate this effect. In this paper, we present the code \prospect{} for computing profile likelihoods in cosmology. We showcase the code using a phenomenological model for converting dark matter into dark radiation that suffers from large volume effects and prior dependence. \prospect{} is compatible with both \cobaya{} and \montepython{}, and is publicly available at \url{https://github.com/AarhusCosmology/prospect\_public}.
\end{abstract}

\begin{keywords}
	cosmological parameters -- cosmic background radiation -- methods: numerical -- methods: statistical
\end{keywords}

\section{Introduction}

Over the past two decades, Bayesian statistics has become the standard tool for most cosmological data analysis. The main reason for this is that Bayesian parameter inference can be carried out efficiently through a single sampling of the full posterior distribution from which all marginalised posterior distributions can be obtained by inexpensive integrations. These integrations, however, require the specification of an integration measure, the prior probability distribution (henceforth \textit{prior})~\citep{bda}. Consequently, the final result may be susceptible to changes in assumed priors, leading to a possible issue with the interpretations thereof. Although the usual uniform priors are non-informative when the likelihood functions are Gaussian in the model parameters~\citep{bda}, as is the case for the standard $\Lambda$CDM model analysed with CMB data from Planck~\citep{Planck:2013nga, Planck:2018vyg}, these issues have received recent interest in a diverse range of cosmological analyses. Examples of these include the prior dependence of Bayesian constraints on the neutrino mass~\citep{Gariazzo:2022ahe} and the effective sterile neutrino mass~\citep{Planck:2018vyg}, the dependence on priors on the parameters of $\Lambda$CDM extensions such as decaying dark matter~\citep{Holm:2022eqq, Holm:2022kkd} and early dark energy~\citep{Herold:2021ksg, Herold:2022iib, Cruz:2023cxy} as well as the dependence on priors on expansion parameters in the effective field theory of large scale structure analyses of BOSS and eBOSS data~\citep{Carrilho:2022mon, Simon:2022lde, Donald-McCann:2023kpx, Zhao:2023ebp, Moretti:2023drg, Holm:2023laa, Gsponer:2023wpm}.

Frequentist statistics, on the other hand, has no notion of prior probability, and is therefore inherently independent of prior choices. Hence, to disentangle the impact of prior effects on results of cosmological data analyses, it has recently been suggested to complement the Bayesian analyses with frequentist results using profile likelihoods~\citep{Herold:2021ksg, Reeves:2022aoi, Simon:2022lde, Holm:2022kkd, Herold:2022iib, Nygaard:2023cus, Holm:2023laa}. Since the latter is obtained from the full-dimensional likelihood function by maximising over, rather than integrating, subsets of the model parameters, it does not require the definition of a prior as an integration measure (accordingly, it carries a different interpretation than the marginalised posterior distributions).

A number of numerical packages exist for the Bayesian cosmological inference, typically consisting of a Markov-chain Monte Carlo (MCMC) engine like \montepython{}~\citep{Audren:2012wb, Brinckmann:2018cvx} or \cobaya{}~\citep{cobaya_code, cobaya_paper} which make calls to an Einstein--Boltzmann solver such as \CAMB{}~\citep{Lewis:1999bs} or \CLASS{}~\citep{Diego_Blas_2011}, as well as to likelihood functions associated with various cosmological data sets. On the other hand, there does not currently exist a generally applicable tool for constructing profile likelihoods. Hence, in this paper we present \prospect{}, a new numerical tool for cosmological profile likelihoods. \prospect{} computes profile likelihoods after a Bayesian analysis has been carried out by interfacing either \cobaya{} or \montepython{}. It is designed to be user-friendly, and employs an efficient gradient-free optimisation procedure based on simulated annealing with adaptive step size. 
\prospect{} is publicly available upon the release of this paper\footnote{\url{https://github.com/AarhusCosmology/prospect\_public}}. 

The rest of the paper is organised as follows. In section~\ref{sec:prior}, we briefly outline the usefulness of the profile likelihood, and in section~\ref{sec:2} we discuss the strategy behind the numerical computation of profile likelihoods in \prospect{}. In section~\ref{sec:3} we provide an profile likelihood analysis with \prospect{} in the form of a phenomenological model for converting dark matter to dark radiation. Finally, we provide our conclusions in section~\ref{sec:conclusion}. In Appendix~\ref{sec:prospect} we discuss a few technical aspects of the code and in appendix~\ref{sec:apB} we provide some additional figures and results on the model discussed in section~\ref{sec:3}.

\section{Prior effects in Bayesian inference}\label{sec:prior}

In Bayesian statistics, the observed data is assumed fixed with zero uncertainty, whereas the model parameters (including nuisance parameters that represent the experimental uncertainty) are stochastic. The central object in Bayesian inference is the probability distribution over the latter, i.e. the posterior probability distribution $P(\bm{\theta})$ (henceforth \textit{posterior}) over a vector of model parameters $\bm{\theta}$. The posterior is defined from the likelihood function $L(\bm{\theta})$ as 
\begin{align}\label{eq:bayes_theorem}
	P(\bm{\theta}) = \frac{L(\bm{\theta}) \pi(\bm{\theta})}{\protect{\int L(\bm{\theta})} \pi(\bm{\theta}) \d \bm{\theta}},
\end{align}
where the prior probability distribution $\pi (\bm{\theta})$ can be seen as the chosen integration measure on parameter space. The posterior is commonly approximated numerically by sampling from it using Markov Chain Monte Carlo (MCMC). In order to construct credible intervals for a parameter of interest $\theta_j$, the posterior is integrated over the other $N-1$ parameters. This proces is called marginalisation, and the outcome is the one-dimensional marginalised posterior distribution,
\begin{align}\label{eq:margpost}
	P(\theta_j) = \protect\int P(\bm{\theta}) \prod_{i\neq j} d\theta_i  \, .
\end{align}
Evidently, credible intervals for the parameter $\theta_j$ may be influenced by the prior $\pi (\bm{\theta})$ both through the definition of the posterior probability~\eqref{eq:bayes_theorem} and through the marginalisation procedure~\eqref{eq:margpost}. In the former case, the priors are said to be informative, and in the latter case, the impact of the prior is called a volume effect. In the following, we discuss the interpretations of each case.

\textbf{Informative priors.} 
Priors can have different motivations. For example, they can represent the uncertainty of an experimental nuisance parameter as derived from simulations, or the theoretical prejudice tied to the model, e.g. naturalness arguments for dimensionless parameters in a Lagrangian. In such well-motivated cases, the ability to impose a prior can be seen as an advantage of the Bayesian analysis. However, when there are no particularly well-motivated priors, the best choice may not be clear. In cosmological parameter inference, it is common to assign uniform priors when no prior information is available, but due to the transformation of probability densities, a uniform prior in one parametrisation will generally be non-uniform in some other parametrisation; for example, a uniform prior in the parameter $\log_{10} \theta$ is equivalent to assigning the prior $\pi (\theta) \propto \theta^{-1}$ on $\theta$. Thus, uniform priors may still be informative. The Jeffrey's prior~\citep{bda} is a choice of prior that is the theoretically \textit{least informative}, and has been recently been employed to minimise the impact of the prior choice on inference results~\citep{Hadzhiyska:2023wae, Donald-McCann:2023kpx, Gsponer:2023wpm}. However, this prior is often non-trivial to compute without knowing the full likelihood functions, and is not guaranteed to completely mitigate the prior effects (it simply minimises them). In a given analysis, then, it is difficult to assess to what extent the chosen priors are informative.

\textbf{Volume effects.}
Looking at equation~\eqref{eq:margpost}, there are two scenarios that can lead to the same large value of $P(\theta_j)$: either the likelihood can be large but in a small region of parameter space, or the prior supports a large volume where the likelihood $L\left(\vec{\theta}\right)$ is modest. As has recently been pointed out in the literature~\citep{Herold:2021ksg, Holm:2022kkd}, this effect is particularly important for many extensions of the $\Lambda\text{CDM}$ model, since these models have a control parameter $f$ which becomes unconstrained in the $\Lambda\text{CDM}$ limit, leading to a large volume of high likelihood which emphasises the $\Lambda$CDM limit of parameter space after marginalisation. Due to this, volume effects can be interpreted as a natural way of penalising additional degrees of freedom in Bayesian model comparison. However, the nature of the volume effect is strongly dependent on the choice of parametrisation; for example, parametrising the aforementioned control parameter as $\log f$ instead of simply $f$ can lead to a much stronger penalisation of the other parameters of the $\Lambda$CDM extension. Volume effects can occasionally be mitigated by reparametrising the model in terms of parameters that have a direct (and maybe even linear) impact on the relevant observables. For example, if $\theta$ is a model parameter that is unconstrained in the limit $f\rightarrow 0$ of the control parameter $f$, one could reparametrise the extension as $(f, \theta) \rightarrow (f, f\theta)$, where, since $f\theta$ will not be unconstrained as $f\rightarrow 0$, one expects no significant volume effect. Unfortunately, for more advanced $\Lambda$CDM extensions, it is not clear what the proper reparametrisation is, and even though such reparametrisations alleviate the volume effects, one risks creating informative priors cf. the discussion above.


In sections~\ref{sec:volume_effects} and~\ref{sec:prior_effects}, we illustrate these two effects in a specific cosmological model. As described, these two effects may well be desireable insofar as they are well-motivated. However, when a practitioner does not have a well-argued prior and simply wishes to get the least prior dependent result, it can be difficult to assess to what extent the resulting constraints are dominated by the two effects just described. By comparing Bayesian results to the corresponding frequentist results, one sees exactly what the influence of the chosen priors are. It is in this sense that the Bayesian and frequentist approaches are complementary. 

\section{Constructing the profile likelihood} \label{sec:2}
In a frequentist context, there is, a priori, no well-defined integration measure over the parameter space, so the most commonly used alternative to the marginalisation procedure~\eqref{eq:margpost} is maximisation, which leads to the \textit{profile likelihood.} It is defined for a parameter $\thetajfixed$ as
\begin{align}
	L(\thetajfixed) &= \max_{\left\{\vec{\theta} | \theta_j=\thetajfixed \right\}}  L\left(\vec{\theta}\right) \,,
\end{align}
where the maximisation runs over the $N-1$ remaining parameters.

While the profile likelihood is occasionally used to construct approximate confidence intervals~\citep{pawitan}, inferences made from it are biased inasmuch as the maximum likelihood estimates of the auxilliary parameters $\hat{\theta_i}, i\neq j$ differ from their true values. Although the latter coincide in the large sample limit, the error made is difficult to assess quantitatively. Recently, reference~\cite{Herold:2024enb} investigated the extent to which this limit is satisfied for a series of extensions to the $\Lambda$CDM model, and found that it is not adequately satisfied even for moderately non-trivial extensions to the $\Lambda$CDM model. For this reason, we refrain from computing confidence intervals directly from the profile likelihood. Instead, the purpose of employing the profile likelihood is to assess the influence of the prior effects described in section~\ref{sec:prior}. In particular, since it is a maximum likelihood estimate in the reduced parameter space, it is invariant under reparametrisations of the latter. Furthermore, it is clearly prior independent. Thus, the profile likelihood is useful to gauge the impact of the prior effects when compared to a corresponding Bayesian analysis.

While profile likelihoods were used to some extent in early cosmological parameter analyses, they were eventually phased out in favour of Bayesian marginalisation because they are numerically expensive. For a typical $6 \lesssim N \lesssim 10$ dimensional cosmological parameter space, each parameter profile evaluated at $M$ points amounts to carrying out $M$ numerical optimisations over all other parameters---an operation which, unlike integration, cannot be reused for other parameters. Therefore, even though each optimisation is cheap, calculating profile likelihoods for many parameters quickly becomes expensive. Nevertheless, profile likelihoods have recently had a renaissance in cosmological data analysis because it has been realised that in many cases Bayesian inference can be dominated by prior effects.

In addition to the large amount of optimisations required, numerical noise from the Einstein--Boltzmann solver leads to noise in the likelihood landscape which in turn makes it difficult to obtain stable numerical gradients. This severely restricts the usefulness of gradient-based optimisation algorithms. Nonetheless,~\cite{Planck:2013nga} showed that it was possible to do gradient-based optimisation of the Planck likelihood in the $\Lambda\text{CDM}$-model if the precision settings of the Einstein--Boltzmann solver were increased significantly. Another promising remedy is to emulate the output of the Einstein--Boltzmann code (e.g.~\cite{SpurioMancini:2021ppk, Nygaard:2022wri, Gunther:2022pto, LINNA, Gammal:2022eob, Bonici:2023xjk, Gunther:2023xhh}), which gives fast gradients with minimal noise. In particular,~\cite{Nygaard:2023cus} uses gradients from a neural network emulator to employ an effective gradient-based optimisation algorithm to construct profile likelihoods for cosmological inference. Another approach is to build an Einstein--Boltzmann code in a framework that permits auto-differentiation, see e.g.~\cite{zack_li_2023_10065126,Hahn:2023nvb} for two such implementations in \texttt{Julia}. Although interesting, these efforts are still in the early stages, and time will tell if a mature and competitive code will emerge which can be used for CMB analyses. 

Given these observations, stochastic gradient-free algorithms emerge as the most generally applicable optimisation method in cosmological inference. One of the most succesful of such algorithms is the method of \textit{simulated annealing}~\citep{Kirkpatrick:1983zz}, first used in cosmology by~\cite{Knox:1995dq} and~\cite{Hannestad:2000wx}, but which has recently been employed on several occasions in cosmological inference, especially to find global maximum likelihood estimates as a supplement to the usual Bayesian analysis (e.g.~\cite{Schoneberg:2021qvd} as well as~\cite{Herold:2021ksg, Reeves:2022aoi, Herold:2022iib, Holm:2022kkd, Cruz:2023cxy, Holm:2023laa, Goldstein:2023gnw, Efstathiou:2023fbn} in the context of profile likelihoods). Given the success of the algorithm, \prospect{} employs a modified version of simulated annealing in its optimisation. In this section, we describe and discuss the adaptive simulated annealing algorithm implemented in \prospect{}, as well as the initialisation of the optimisation using an MCMC.

\subsection{Simulated annealing}\label{sec:simanneal}
Simulated annealing is a gradient-free stochastic optimisation algorithm based on the behaviour of thermodynamic systems cooling down~\citep{Kirkpatrick:1983zz}. In practice, it works by iteratively running an MCMC~\citep{10.1063/1.1699114} chain on the parameter space with a Gaussian proposal distribution centered on the current position in parameter space $\vec{\theta}$ and the modified acceptance probability,
\begin{align}\label{eq:acceptance_probability}
	p = - \frac{\log L (\vec{\theta}) - \log L(\vec{\theta}_\mathrm{proposed})}{T},
\end{align}
where the parameter $T$, referred to as the \textit{temperature}, is decreased after each iteration. An equivalent interpretation is that the MCMC chain runs with the usual acceptance probability $p=\log \tilde{L}(\vec{\theta}_\mathrm{proposed}) - \log \tilde{L}(\vec{\theta})$ but on the modified likelihood surface $\tilde{L}\equiv L^{1/T}$. From this view, as the temperature decreases, peak structures in the likelihood surface are enhanced, leading to an increasing probability of the MCMC residing in an optima. To remedy the stochastic nature of the algorithm, \prospect{} carries out $m$ independent optimisations at each point in the profile likelihoods, where $m$ is a user-defined input parameter, typically around $2$--$3$, and taking the best value obtained among these. 

Simulated annealing generally works well against noisy likelihoods with many local optima. The main difficulty in applying simulated annealing is the tuning of its hyperparameters that control the temperature and the proposal density for the MCMC at each iteration. Although it is only provably convergent to the global optimum when the temperature decreases logarithmically with iteration number~\citep{SA_proof}, this choice leads to slow convergence in many practical scenarios~\citep{INGBER199329, ingber2000adaptive}. It is therefore common to decrease the temperature faster than logarithmically, e.g.\ exponentially\footnote{When the temperature decreases exponentially, the correct name of the algorithm is \textit{simulated quenching}~\citep{ingber2000adaptive}; however, in this paper, we stick to the nomenclature of simulated annealing for simplicity.}, at the cost of the theoretical guarantee of global convergence. In practice, this works well when starting the optimisation from the basin of the global optimum, which is most often the case with the initialisation procedure to be described shortly. Given these considerations, \prospect{} employs an exponentially decaying temperature schedule by default and uses a Gaussian proposal distribution $\mathcal{N}(\vec{\theta}, S\Sigma)$, where $\vec{\theta}$ is the current position in parameter space, $\Sigma$ is the covariance matrix and the \textit{step size} $S$ is a real number that scales the covariance.

Whereas the proposal covariance matrix $\Sigma$ defines the relative step lengths between different directions in parameter space, the step size $S$ defines a global scaling of the step length in all directions. At any given temperature, a too large step size leads to a small acceptance rate, while a too small step size gives an inefficient exploration of parameter space. It is therefore important to balance the step size by varying it simultaneously with the temperature.

These heuristics suggest a tuning of the step size such that the acceptance rate $A$, i.e.\ the ratio of accepted to attempted steps in a given iteration, is neither large nor small. In \prospect{}, we accomplish this by modifying the step size $S_i$ at the $i$'th iteration according to
\begin{align} \label{eq:adaptive_step_size}
	S_{i+1} = \begin{cases}
		(1 - m)S_i, \qquad \text{if }A_i < A_\mathrm{t}, \\
		(1 + m)S_i, \qquad \text{if }A_i > A_\mathrm{t},
	\end{cases}
\end{align}
where the multiplication rate $m$ and the target acceptance rate $A_\mathrm{t}$ are user-defined parameters. The logic behind the two cases above is the following: If the current acceptance rate $A_i$ is smaller (larger) than the target acceptance rate $A_\mathrm{t}$, decreasing (increasing) the step size will result in smaller (larger) likelihood differences between the proposal and current positions, possibly giving a larger (smaller) acceptance rate in the next iteration. In practice, of course, the acceptance rate in the next iteration depends more seriously on the shape of the likelihood surface. For example, there is a natural decline in the acceptance rate as the temperature is decreased. It is therefore important that the multiplication rate $m$ is large enough that the step size adaptations can overcome the changing temperature. Additionally, one would naturally expect smaller acceptance rates near a global minimum due to the form of the acceptance probability~\eqref{eq:acceptance_probability}. Nonetheless, we find that a step size tuning of this form is advantageous both because it relieves the user of having to provide a concrete schedule and because it greatly outperforms most monotonic step size schedules. We have found $m\approx 0.5$ and $A_\mathrm{t} \approx 0.2$ to give good results.

\begin{figure*}
	\centering
	\includegraphics[width=0.8\textwidth]{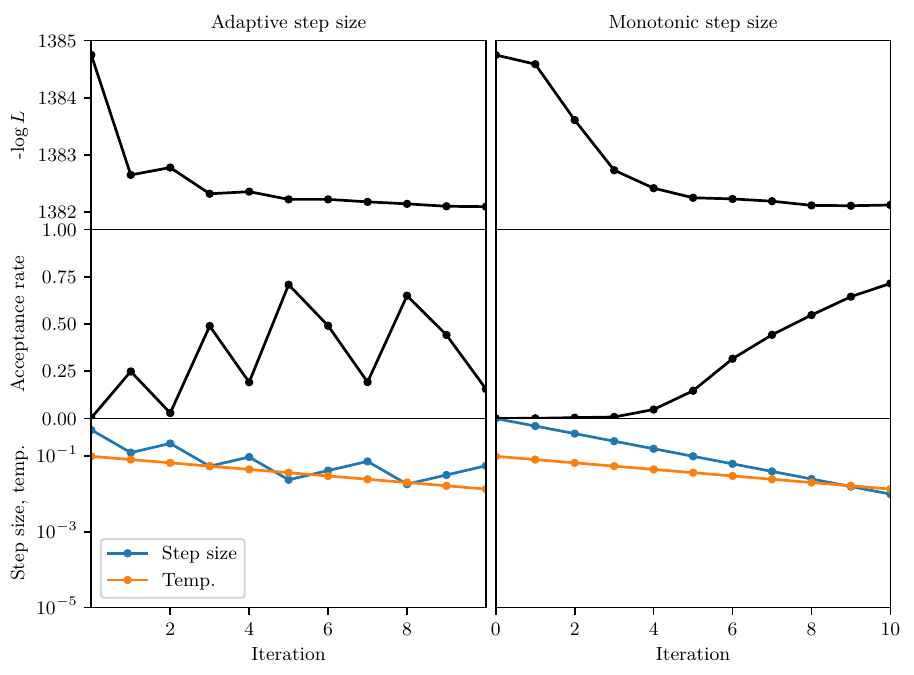}
	\caption{Example statistics of a single simulated annealing optimisation of the decaying cold dark matter model~\protect\citep{Audren:2014bca, Poulin:2016nat, Nygaard:2020sow}, subject to Planck high-$\ell$ TTTEEE and low-$\ell$ TT and EE data, with fixed decay constant $\Gamma_\mathrm{DCDM} = 10^{8.077} \text{ km s}^{-1}\text{ Mpc}^{-1}$, computed using the \connect{} emulator of~\protect\cite{Nygaard:2023cus}. The left and right columns represent optimisations using the adaptive method of this section and a monotonic, exponentially decreasing step size, respectively. \textit{Top row:} The negative logarithm of the likelihood function to optimise. \textit{Middle row:} The acceptance rate (ratio of accepted to attempted MCMC steps). \textit{Bottom row:} The step sizes and temperatures. Evidently, the adaptive algorithm, in addition to relieving the user of having to choose a particular schedule, also performs better, and avoids several common pitfalls of monotonically decreasing step size schedules.}
	\label{fig:schedules}
\end{figure*}

To illustrate some of these considerations, figure~\ref{fig:schedules} shows several statistics of the optimisation algorithm as a function of the iteration number. The likelihood function optimised here is an extension of the $\Lambda$CDM model where a fraction of the cold dark matter decays to dark radiation on cosmological timescales (e.g.~\cite{Audren:2014bca, Poulin:2016nat, Nygaard:2020sow}), subject to Planck high-$\ell$ TTTEEE and low-$\ell$ TT and EE data~\citep{Planck:2018vyg}. This model was recently studied with profile likelihoods in~\cite{Nygaard:2023cus} and~\cite{Holm:2022kkd}, and the results presented here are obtained with the \textsc{connect} neural network from~\cite{Nygaard:2023cus}. The single optimisation shown in figure~\ref{fig:schedules} has the fixed decay constant of cold dark matter $\Gamma_\mathrm{DCDM} = 10^{8.077} \text{ km s}^{-1}\text{ Mpc}^{-1}$. The first row shows the descend of the negative logarithm of the likelihood value as the iterations pass; the second row shows the acceptance rate of the simulated annealing in each iteration, and the last row shows the temperature (red) and step size (blue) at each iteration. The left column represents an optimisation with the adaptive step size procedure described in this section (with $m=0.75$ and $A_\mathrm{t}=0.2$), whereas the right column represents a typical monotically decreasing step size schedule. The temperature schedule, decreasing exponentially, is the same in both columns.

It is clear from the figure that the step size prescription~\eqref{eq:adaptive_step_size} achieves the goal of keeping the acceptance rate on the order of the target acceptance rate. In this particular case, both optimisations start out with negligible acceptance rate due to a large step size. In response, the adaptive method quickly decreases the step size, giving a larger acceptance rate, leading to more accepted steps and consequently better progress toward the minimum in the first iterations. At later iterations, the monotonically decreasing step size becomes very small compared to the temperature, leading to an asymptotic saturation of the acceptance rate of $1.0$ and no change in the likelihood since the parameter space steps taken are vanishingly small. This is a common issue of the monotonically decreasing step size schedules, which usually has to be accommodated by \textit{reannealing}, i.e.\ by discontinuously increasing the temperature and step sizes back up to high values and commencing a new simulated annealing from there. Although this is sometimes also needed for the adaptive procedure, the latter is more robust against such limiting behaviours since it keeps the acceptance rate from saturating. 

Finally, the run with the monotonically decreasing step size schedule is also strongly dependent on the particular initial value and decay constant chosen, thus relying directly on the experience and intuition of the user. Moreover, the optimal schedule may differ greatly across the different fixed points in the profile likelihood. Ultimately, then, the adaptive step size procedure, although simple, is both more efficient and robust than monotonic schedules and crucially relieves the user of having to choose a specific schedule.

\subsection{Initialisation from MCMC}\label{sec:initialisation}
The assumed use-case of \prospect{} is that the user has finished a Bayesian analysis with an MCMC and now desires to construct profile likelihoods to compare with. Although MCMC chains are poor optimisers~\citep{Hamann_2012}, approximate profile likelihoods can be obtained from them by binning. This was done in~\cite{Gomez-Valent:2022hkb} as a fast test of the severity of volume effects\footnote{This estimate can be noisy when there are too few samples in the MCMC. In order to assess the impact of marginalisation, the mean likelihood value in such bins has occasionally been employed instead (e.g.~\cite{Ivanov:2020ril, McDonough:2023qcu}).}. \prospect{} uses this principle to estimate a profile likelihood to start the optimisations from. 
Before starting the optimisation of a given point $\thetajfixed$, \prospect{} selects the fraction $\xi$ of points in the Markov-chain which has $\theta_j$ closest to $\thetajfixed$, and takes the starting point for the simulated annealing to be the highest likelihood point in this subsample with $\theta_j \rightarrow \thetajfixed$. Typically, $\xi$ is on the order of $0.1$, but there is a trade-off in choosing $\xi$, since a lower value leads to more local information at the expense of more sampling noise.

\begin{figure*}
	\centering
	\includegraphics[width=0.8\textwidth]{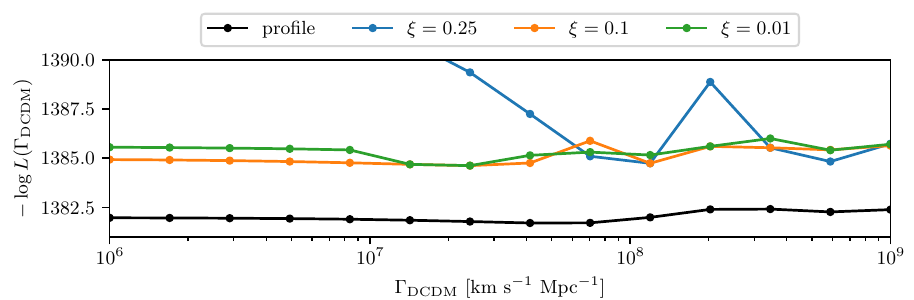}
	\caption{Profile likelihood (black) computed with \prospect{} for the decaying cold dark matter model of figure~\protect\ref{fig:schedules} using Planck high-$\ell$ TTTEEE, low-$\ell$ TT and EE data~\protect\citep{Planck:2018vyg}, along with estimates of this profile using a fully converged MCMC of the same model with different bin fractions $\xi$, as explained in the text. The estimates using $\xi=0.1$ and $\xi=0.01$ are only $\Delta \log L \approx 5$ above the fully converged profile, and hence are good initial points for the simulated annealing optimisation.}
	\label{fig:initialisation_profile}
\end{figure*}
To study the effects of different choices of the bin fraction $\xi$, figure~\ref{fig:initialisation_profile} shows a converged profile likelihood of the decay constant $\Gamma_\mathrm{DCDM}$ of the decaying cold dark matter model of figure~\ref{fig:schedules} using Planck high-$\ell$ TTTEEE, low-$\ell$ TT and EE data~\citep{Planck:2018vyg}, as well as the profile likelihood estimates constructed from a fully converged MCMC of the model (both again computed using \textsc{connect}~\citep{Nygaard:2022wri, Nygaard:2023cus}), for three different values of the binning fraction $\xi$. Although the converged profile likelihood seems flat due to the large scale on the second axis, the negative logarithm of the likelihood has a shallow well with its global bestfit around $\Gamma_\mathrm{DCDM}\approx 7\times 10^{7} \text{ km s}^{-1} \text{ Mpc}^{-1}$. 
In the figure, we also observe that the estimated profiles using the values $\xi=0.1$ and $\xi=0.01$ are almost identical, whereas the profile estimated using $\xi=0.25$ is somewhat worse. 
%
The optimal choice of the bin fraction $\xi$ will depend on the size of the MCMC chain one is starting from: If it is particularly large, smaller $\xi$ should give more accurate profile likelihood estimates, and vice-versa for small MCMC chains. Currently, \prospect{} users must supply a value for $\xi$, but as future work, we plan on experimenting with adaptive schemes for automatically choosing $\xi$, for example by allowing different $\xi$ in each bin and finding the optimal starting point by optimising over $\xi$. 
In the end, however, the quality of the initial profile does not depend strongly on the exact choice of $\xi$, as figure~\ref{fig:initialisation_profile} shows, and we have found $\xi=0.1$ to work well for all cases studied in this paper.



\begin{figure*}
	\centering
	\includegraphics[width=0.8\textwidth]{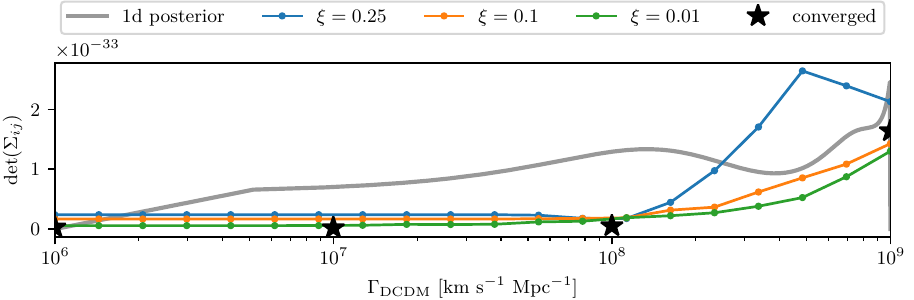}
	\caption{\label{fig:initialisation_covmat} Determinants of parameter covariance matrices of the decaying cold dark matter model of figures~\ref{fig:schedules}--\ref{fig:initialisation_profile}, for fixed values of the decay constant $\Gamma_\mathrm{DCDM}$, for fully converged MCMCs (black) and as estimated from an MCMC with varying $\Gamma_\mathrm{DCDM}$ using the binning procedure described in the text. Superimposed on the figure is the one-dimensional marginalised posterior distribution of $\Gamma_\mathrm{DCDM}$, arbitrarily normalised. The latter is influenced by a strong volume effect at the upper prior bound of its MCMC, leading to a widening of the posterior and hence an increase in the fixed-parameter covariance matrix determinants at large $\Gamma_\mathrm{DCDM}$. Importantly, the covariance matrix estimates all capture this effect, and are consequently expected to lead to better simulated annealing optimisation than if the global covariance matrix were used.}
\end{figure*}
In addition to constructing starting points for the optimisations, \prospect{} also constructs local covariance matrices for the Gaussian proposal distributions from the MCMC at each point in the profile. This is especially useful since, for example, the global covariance matrix of the MCMC is prone to volume effects which can distort the relative length scales in different dimensions. Figure~\ref{fig:initialisation_covmat} illustrates how these fixed-parameter covariance matrices differ as a function of the values of the profile likelihood parameter, again with the decaying cold dark matter model of the last sections as an example. Since it is difficult to visually compare the full $29\times 29$-dimensional covariance matrices, we show the determinant of the covariance matrices for different fixed values of the dark matter decay constant $\Gamma_\mathrm{DCDM}$. Although a heavily summarising statistic, the determinant of the covariance matrix is a measure of the curvature of the posterior probability distribution around its maximum a-posteriori estimate, and hence is a measure of the product of the length scales along different dimensions in parameter space. Nevertheless, we are not interested in the exact interpretation of this metric, but rather its variation across different fixed values of the profile parameter $\Gamma_\mathrm{DCDM}$. 

Hence, in figure~\ref{fig:initialisation_covmat}, we show the value of the covariance matrix determinant for a set of four fully converged MCMC runs at the fixed points $\Gamma_\mathrm{DCDM} \text{ km}^{-1} \text{ s Mpc} \in \{10^6, 10^7, 10^8, 10^9\}$ as well as its estimated values for different choices of the bin fraction $\xi$. Superimposed on the figure is the one-dimensional marginalised posterior distribution of $\Gamma_\mathrm{DCDM}$ at arbitrary normalisation. This posterior should not be compared directly to the determinant curves, but is there to indicate a well-understood volume effect~\citep{Holm:2022kkd} by its peaking toward the upper prior bound of its MCMC at $\Gamma_\mathrm{DCDM}=10^9 \text{ km s}^{-1}\text{ Mpc}^{-1}$. The volume effect widens the posteriors around large values of $\Gamma_\mathrm{DCDM}$, which increases the covariance in the widened dimensions, leading to a larger value of the determinant of the fixed-parameter covariance matrices. This effect is seen directly in figure~\ref{fig:initialisation_covmat}, and underlines the importance of not relying on a global covariance matrix for simulated annealing when there are serious volume effects in the model. 
Seemingly, the exact value of the bin fraction $\xi$ does not play a large role, and as discussed above, although clever tunings of $\xi$ can be introduced, we have found the value $\xi=0.1$ to work well for all the cases studied in this paper.

\section{A worked example: Conversion of dark matter to dark radiation} \label{sec:3}
In this section, we present results on an extension to the $\Lambda$CDM cosmology that enables a parametrised transition of a fraction of the cold dark matter to dark radiation~\citep{Bringmann:2018jpr, McCarthy:2022gok}. As we will see, the three added parameters display serious volume effects since $\Lambda$CDM is recovered in several limiting cases. Consequently, the model exemplifies a case where Bayesian and frequentist inferences may not agree, thus proving the importance of combining them for a fully nuanced analysis.

\subsection{Model description}\label{sec:model}
The model we study introduces a parametrised conversion of part of the dark matter to dark radiation (henceforth the DMDR model). It is defined by a phenomenological parametrisation of the homogeneous cold dark matter energy density as~\citep{Bringmann:2018jpr}
\begin{align}\label{eq:cdm}
	\rho_\mathrm{CDM} (a) = \frac{\rho_\mathrm{CDM}^0}{a^3} \left( 1 + \zeta \left( \frac{1 - a^\kappa}{1 + (a/a_\mathrm{t})^\kappa} \right) \right),
\end{align}
where $a$ denotes the scale factor (taking $a=1$ today), $\rho_\mathrm{CDM}^0$ the cold dark matter density today, and we have introduced three model parameters: $1+\zeta$ describes the fraction of the dark matter that transitions, $\kappa$ controls the shape of the decay of the dark matter density and $a_\mathrm{t}$ denotes the scale factor around which the transition is centered. In the limit of large $\kappa$, the transition is instant at $a=a_\mathrm{t}$, getting more stretched in time as $\kappa\rightarrow 0$. The special case $\kappa=1$ corresponds to Sommerfeld-enhanced annihilations, and $\kappa=2$ approximates an exponentially decaying dark matter sector~\citep{Bringmann:2018jpr}.

The energy density~\eqref{eq:cdm} defines the energy density of the dark radiation $\rho_\mathrm{DR}$, into which it transitions, through energy conservation~\citep{Bringmann:2018jpr}
\begin{align}\label{eq:energy_cons}
	a^{-3} \dv{(a^3\rho_\mathrm{CDM})}{a} = -a^{-4} \dv{(a^4 \rho_\mathrm{DR})}{a} \equiv \mathcal{Q},
\end{align}
where $\mathcal{Q}>0$ describes the energy transfer between the species. The parametrisation~\eqref{eq:cdm} does not uniquely specify the first order perturbation around the homogeneous solution. Here, we adopt the choice of~\cite{Bringmann:2018jpr, McCarthy:2022gok} to set the linear perturbation of the energy transfer proportional to the dark matter density, $\delta \mathcal{Q} \equiv \mathcal{Q} \delta_\mathrm{CDM}$, as is the case for decaying dark matter~\citep{Audren:2014bca}\footnote{In fact, this choice makes the first order equations of the DMDR model equivalent to a decaying dark matter model with the time-dependent decay constant $\Gamma(a)=\mathcal{Q}(a)/\rho_\mathrm{CDM}(a)$~\citep{McCarthy:2022gok}.}. Covariant conservation of energy-momentum then gives the equations for the linear perturbations of the dark matter energy density in synchronous gauge as
\begin{align}
	\delta_\mathrm{CDM}' + \frac{1}{2}h' = 0,
\end{align}
in the usual notation of~\cite{Ma:1995ey}, where $h$ denotes the trace of the spatial part of the perturbed Friedmann–Lemaître–Robertson–Walker metric in synchronous gauge and the prime denotes conformal time derivatives. The velocity divergence vanishes by definition in the comoving synchronous gauge~\citep{Audren:2014bca}, and higher moment perturbations vanish by the cold dark matter assumption~\citep{Ma:1995ey}. The moments $\Psi_{\mathrm{DR},\ell}$ of the perturbed dark radiation distribution function are summarised by the momentum-integrated moments 
\begin{align}\label{eq:dr_moment_defn}
	F_{\mathrm{DR},\ell} (a, k) \equiv r_\mathrm{DR}(a) \frac{\int \d p \ p^3 f_\mathrm{DR}^{(0)}(a, p) \Psi_{\mathrm{DR},\ell}(a,k,p) }{\int \d p \ p^3 f_\mathrm{DR}^{(0)} (a,p)},
\end{align}
with $r_\mathrm{DR}(a) \equiv a^4 \rho_\mathrm{DR} (a) /\rho_\mathrm{crit,0}$, where $\rho_\mathrm{crit,0}$ is the critical energy density of the Universe today and  $f_\mathrm{DR}^{(0)}$ is the homogeneous part of the dark radiation distribution function.
The time evolution of the moments~\eqref{eq:dr_moment_defn} is then given by covariant conservation of energy-momentum as~\citep{McCarthy:2022gok}
\begin{align}
	F_{\mathrm{DR},0}' &=  -k F_{\mathrm{DR},1} -  \frac{4h'}{6} r_\mathrm{DR} + r_\mathrm{DR}' \delta_\mathrm{CDM} \,, \\
	F_{\mathrm{DR},1}' &= \frac{k}{3} F_{\mathrm{DR},0} - \frac{2k}{3} F_{\mathrm{DR},2} \,, \\
	F_{\mathrm{DR},\ell}' &= \frac{k}{2\ell + 1} \left( \ell F_{\mathrm{DR}, \ell - 1} - (\ell + 1) F_{\mathrm{DR}, \ell + 1} \right) \,, \quad \ell \geq 2\,,
\end{align}
in the synchronous gauge.

For transitions that are approximately complete before recombination, the main impact of the model at the homogeneous level is to decrease the sound horizon at last scattering due to the injection of radiation in the early Universe. Since data strongly constrains the angular scale $\theta_s$ of baryon acoustic oscillations, this results in an increased value of the Hubble constant $H_0$. At the perturbed level, the principal imprint on the CMB spectrum is a reduced amount of lensing and an increased integrated Sachs-Wolfe plateau since additional dark energy is required in order to preserve flatness of the Universe as dark matter is converted to the more rapidly diluting dark radiation~\citep{Bringmann:2018jpr}.

The model was initially introduced in~\cite{Bringmann:2018jpr} to test the simple picture of a comovingly constant dark matter density. With a Bayesian analysis, they concluded that CMB data alone excludes the model, whereas late-time data such as the $H_0$ measurement from the Hubble space telescope~\citep{Riess:2016jrr} introduces a $\sim 2\sigma$ ``preference'' for the model due to its prediction of an increased value of $H_0$. However, to combat the strong prior dependence of this preference, the authors of~\cite{Bringmann:2018jpr} conducted an approximate frequentist analysis by approximating the likelihood ratio as the posterior ratio, which yielded only limited preference for the model. More recently,~\cite{McCarthy:2022gok} revisited the model in light of the increasing tensions between local measurements of the Hubble constant and $\sigma_8$, the amplitude of matter fluctuations at $8\,\text{Mpc}/h$ scales, and their values when inferred from CMB data~\citep{Abdalla:2022yfr}. Based on a Bayesian analysis,~\cite{McCarthy:2022gok} concluded that CMB data alone rules out the model as a solution to the tensions, a conclusion which was robust for several different prior choices, although bounds on the DMDR parameters were strongly prior-dependent. Thus, both references agree that the model does not solve the cosmological tensions, but find that priors and parametrisation choices significantly influence the constraints on the DMDR parameters. Hence, in this section, we use the DMDR model as an application of profile likelihoods computed with \prospect{} and demonstrate how the frequentist approach complements the Bayesian analysis in a fruitful way. 

\subsubsection*{Analysis methods and data}
In the following, we compute observables in linear perturbation theory using the model implementation of~\cite{McCarthy:2022gok}\footnote{\url{https://github.com/fmccarthy/class\_DMDR}} in the Einstein-Boltzmann solver \class{}~\citep{Diego_Blas_2011}\footnote{\url{https://github.com/lesgourg/class\_public}}. In this section, we present both Bayesian and frequentist constraints. The former are derived from posteriors sampled with MCMC chains computed using \montepython{}~\citep{Audren:2012wb, Brinckmann:2018cvx}\footnote{\url{https://github.com/brinckmann/montepython\_public}}. We run six MCMC chains for each inference and consider the chains converged when the Gelman-Rubin metric $R-1$ is around $\approx 0.05$~\citep{GELMAN}, although occassionally larger when there are unconstrained parameters. The profile likelihoods are computed using \prospect{}, the novel profile likelihood code published with this paper\footnote{\url{https://github.com/AarhusCosmology/prospect\_public}}. We have started the profile likelihood computation from the MCMCs as described in section~\ref{sec:initialisation}, giving very efficient initial profiles and covariance matrices that improve the optimisation. Furthermore, we use the adaptive step size algorithm with $m=0.5$ and $A_t = 0.2$.

Our baseline data set consists of the following:
\begin{itemize}
	\item Planck 2018~\citep{Planck:2018vyg} high-$\ell$ TTTEEE, low-$\ell$ EE, low-$\ell$ TT and lensing.
	\item Baryon Acoustic Oscillations (BAO) measurements from BOSS DR12~\citep{boss2016}, the main galaxy sample of BOSS DR7~\citep{ross2014} and 6dFGS~\citep{beutler2011}.
	\item Pantheon supernova data~\citep{Pan-STARRS1:2017jku}.
\end{itemize}
Note that we have chosen this baseline to coincide with that of~\cite{McCarthy:2022gok} in order to facilitate comparisons with their results. To assess the effect on cosmological tensions, in section~\ref{sec:h} we additionally include the following data:
\begin{itemize}
	\item A prior on $H_0$ as reported by the SH0ES collaboration~\citep{Riess:2021jrx}.
	\item A prior in $S_8\equiv \sigma_8 (\Omega_\mathrm{M}/0.3)^{0.5}$ from the Dark Energy Survey (DES) year 3 data release~\citep{DES:2021wwk}.
\end{itemize}

\subsection{Volume effects in the DMDR model}\label{sec:volume_effects}
Interestingly, the model recovers $\Lambda$CDM cosmology in several limits of the three model parameters:
\begin{itemize}
	\item $\zeta \rightarrow 0$, i.e.\ in the limit that none of the dark matter undergoes a transition,
	\item $\kappa \rightarrow 0$, in which case the transition is stretched infinitely in time,
	\item $a_\mathrm{t}^\kappa \gg 1$, in which case the transition occurs after today and is too narrow to have affected current cosmology. Furthermore, as $a_\mathrm{t}^\kappa \rightarrow 0$, the model is equivalent to a simple model of additional relativistic degrees of freedom $\Delta N_\mathrm{eff}$\footnote{This limit, also occuring in the simpler decaying dark matter scenario, has been shown to constitute a volume effect in what is the $\zeta$ parameter here~\citep{Holm:2022kkd}.}.
\end{itemize}
In either of these limits, the other two parameters are unconstrained. Since $\Lambda$CDM is known to be a good fit to a large variety of data, these limits correspond to large volumes of high likelihood, suggesting the presence of potentially strong volume effects, increasing the Bayesian affinity for the $\Lambda$CDM regime. Ultimately, this suggests that a profile likelihood analysis, not being affected by volume effects, could provide an illuminating and complementary view on the model.

As an example, we study one of these volume effects in more detail. In the limit $a_\mathrm{t}^\kappa \ll 1$, the decay happens early enough that the decay is complete before the time of recombination, where our most constraining data is taken. The only effect of the model is therefore the injection of dark radiation, which raises the effective number of relativistic species $\Delta N_\mathrm{eff}$. We can construct an analytical expression for the equivalent $\Delta N_\mathrm{eff}$ in this limit as follows. Firstly, energy conservation~\eqref{eq:energy_cons} allows a closed expression of the dark radiation energy density in terms of the Gaussian hypergeometric function~\citep{Bringmann:2018jpr},
\begin{align}
	\rho_\mathrm{DR}(a) = &\zeta \frac{\rho^0_\mathrm{DR}}{a^3} \frac{1+a_\mathrm{t}^\kappa}{a^\kappa + a_\mathrm{t}^\kappa}  \nonumber \\
	&\times \left( (a^\kappa+a_\mathrm{t}^\kappa)  {}_2F_1 \left[1,\frac{1}{\kappa},1+\frac{1}{\kappa},-\left(\frac{a}{a_\mathrm{t}}\right)^\kappa \right]-a_\mathrm{t}^\kappa \right).
\end{align}
We then study the hypergeometric function ${}_2F_1 \left[1,\frac{1}{\kappa},1+\frac{1}{\kappa},-\left(\frac{a}{a_\mathrm{t}}\right)^\kappa \right]$ and note that in the limit $a/a_\mathrm{t} \to \infty$ the leading order term for $0 < \kappa < 1$ is proportional to $(a_\mathrm{t}/a)^\kappa$ which means that the dark radiation density continues to grow at arbitrarily late times and does not map to a fixed $\Delta N_{\rm eff}$ (for $\kappa=1$ the leading term is $\log(a/a_\mathrm{t}) a_\mathrm{t}/a$). For $\kappa > 1$ the leading term is proportional to $a_\mathrm{t}/a$ which means that $ \rho_{\rm DR}$ scales as $a^{-4}$ and can be mapped to a specific $\Delta N_{\rm eff}$. In order to find this values we note 
that we can use the following identity for the hypergeometric function~\citep{abramowitz+stegun}:
\begin{equation}
	_2 F_1(a,b,c,z) = (1-z)^{-b}  {}_2 F_1 (b,c-a,c,z/(z-1)) \,.
\end{equation}
The limit we are interested in is the one in which $a/a_\mathrm{t} \to \infty$ which corresponds to $z \to -\infty$ and therefore to $z/(z-1) \to 1$.
In this particular limit we then find that 
\begin{equation}
	\rho_{\rm DR} \left( \frac{a}{a_\mathrm{t}}\to \infty\right) = \zeta \frac{\rho^0_{\rm CDM} a_\mathrm{t}}{a^4} \frac{\pi \csc(\pi/\kappa)}{\kappa},
\end{equation}
or, equivalently\footnote{Note that, as $\kappa \to 1$, this approximation becomes progressively poorer (i.e.\ requires larger and larger values of $a/a_\mathrm{t}$ to be precise).},
\begin{align}\label{eq:delta_neff}
	\Delta N_\mathrm{eff} = N_\mathrm{eff} \frac{a^4 \rho_\mathrm{DR}  \left( \frac{a}{a_\mathrm{t}}\to \infty\right)}{\rho_\mathrm{\nu}^0} = N_\mathrm{eff} \frac{\rho_\mathrm{CDM}^0}{\rho_\mathrm{\nu}^0} \frac{\pi \csc (\pi / \kappa)}{\kappa} \zeta a_\mathrm{t},
\end{align}
where $\rho_\mathrm{\nu}^0$ is the energy density of neutrinos today. CMB data is only weakly sensitive to $\Delta N_\mathrm{eff}$; for example, the \textit{Planck} collaboration found the $95 \%$ C.L.\ constraint $N_\mathrm{eff}=2.99^{+0.34}_{-0.33}$
using high-$\ell$ TT, TE, EE, low-$\ell$ EE, CMB lensing and BAO data, which is, apart from the inclusion of supernova data, our exact baseline dataset in the data analysis below~\citep{Planck:2018vyg}. Thus, we expect our baseline data to be insensitive to changes on the order of $\Delta N_\mathrm{eff} \lesssim 0.1$. Hence, any combination of $\zeta$, $a_\mathrm{t}$ and $\kappa$ that satisfies this will be indistinguishable from $\Lambda$CDM. By~\eqref{eq:delta_neff}, this defines a region in parameter space where there are large correlations between the three DMDR parameters, giving a large posterior volume, and hence a volume effect, which, as we will see shortly, biases the posterior toward the limit of small $a_\mathrm{t}$. 


In our baseline analysis, we find the globally best-fitting set of cosmological parameters in the DMDR model to have $\zeta = 0.01$, $\log_{10} a_\mathrm{t} = -3.83$ and $\kappa = 3.25$. For these parameters, we have $(a_\mathrm{today}/a_\mathrm{t})^\kappa = 1.8\times 10^{12} \gg 1$, so the model is equivalent to a $\Delta N_\mathrm{eff}$ model with $\Delta N_\mathrm{eff} = 0.035$, as obtained from computing the cosmology in \class{}. As just argued, this is much less than the expected sensitivity of our baseline dataset. In conclusion, the bestfit of the DMDR model with our baseline data is in the $\Lambda$CDM limit. 

\begin{figure*}
	\centering
	\includegraphics[width=0.8\textwidth]{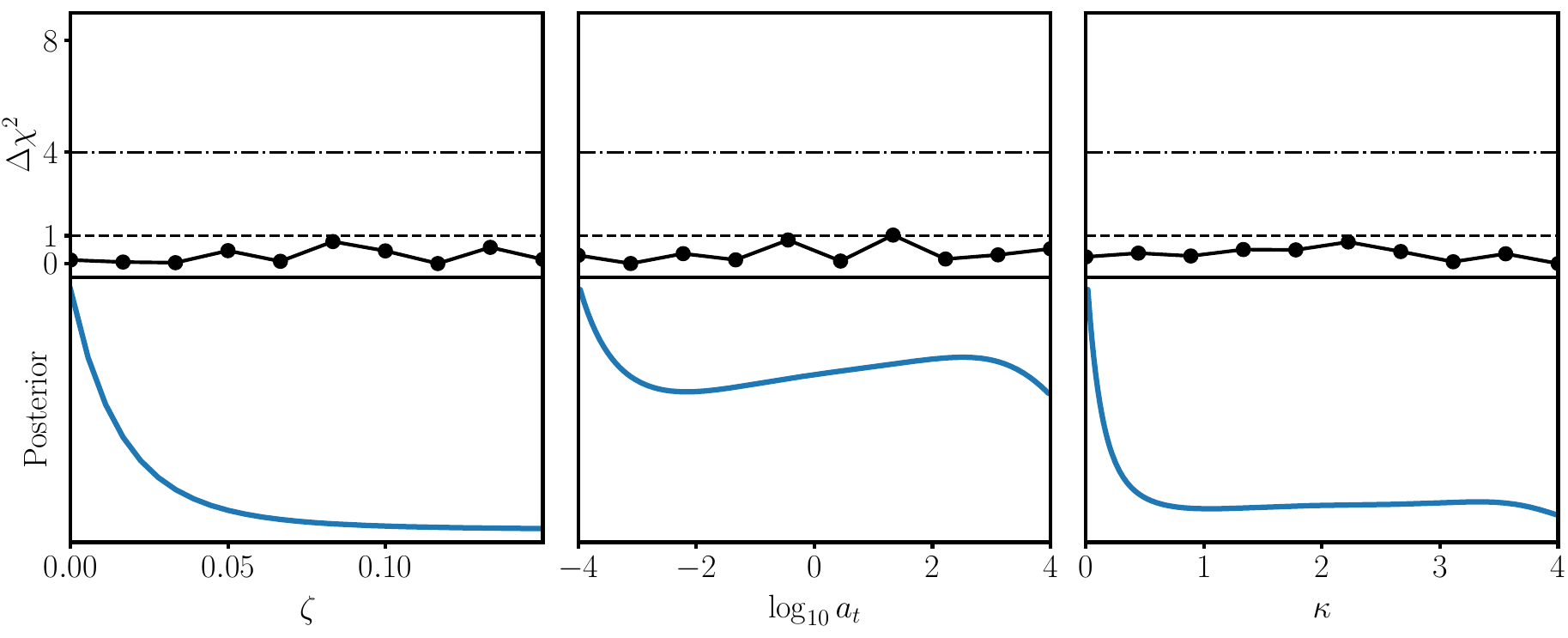}
	\caption{\label{fig:flat_profiles} Constraints on the three parameters of the DMDR model $\zeta$, $\log_{10}a_\mathrm{t}$ and $\kappa$ under the baseline dataset. \textit{Top row:} Profile likelihoods in terms of $\Delta \chi^2$ quantity. \textit{Bottom row:} One-dimensional marginalised posterior distribution from an MCMC. Since the global maximum likelihood estimate of the DMDR model is in the $\Lambda$CDM limit, the profiles of the three parameters are flat. The posteriors, on the other hand, increase toward the lower prior bound due to volume effects.}
\end{figure*}
As a consequence, it is clear, before any inference has been done, that the profile likelihoods of the three DMDR parameters $\zeta$, $\log_{10} a_\mathrm{t}$ and $\kappa$ will be uniform with the $\Lambda$CDM maximum likelihood, since when fixing one of the parameters, it is always possible to obtain a $\Lambda$CDM cosmology by adjusting the other two.

Indeed, this is what we find: Figure~\ref{fig:flat_profiles} shows, in the top panels, profile likelihoods of the three DMDR parameters, computed using \prospect{} with the baseline dataset. Due to the large parameter correlations in the $\Lambda$CDM, the model is difficult to optimise precisely for these fixed parameter values. We find all three profiles to be flat to within $\Delta \chi^2 < 1$ or $\log L < 0.5$, which, is somewhat larger than the usual accuracy of \prospect{} optimisations but still consistent with overall flat profiles, as expected.

The bottom row of figure~\ref{fig:flat_profiles} shows the corresponding one-dimensional marginalised posteriors. Clearly, the posteriors are not flat, but increase toward the lower prior bound of each parameter. Since volume effects are by definition the only difference between profile likelihoods and posteriors with flat priors, this increase of the parameters is solely due to volume effects. The specific volume effects have direct physical interpretations: For $\zeta \rightarrow 0$, no dark matter undergoes a transition, for $a_\mathrm{t} \rightarrow 0$, the decay happens so early that the radiation produced has redshifted away long before recombination, and for $\kappa \rightarrow 0$, the transition is infinitely stretched. 

Hence, this example illustrates how knowing the profile likelihoods allows one to detect volume effects and understand where they come from. In conclusion, the upper bounds one would derive from these posteriors are exclusively driven by volume effects. As discussed in section~\ref{sec:prior}, volume effects may be desirable, but still then, profile likelihoods are useful in identifying them. In the next section, we illustrate an additional subtlety of the posterior that the profile likelihood helps to illuminate.

\subsection{Prior dependence of the Bayesian analysis}\label{sec:prior_effects}
\cite{Bringmann:2018jpr} and~\cite{McCarthy:2022gok} both discussed the effect of different DMDR parameter priors on the results of the Bayesian inference.~\cite{Bringmann:2018jpr} argued that the prior dependence of the Bayesian constraints on $\zeta$ and $a_\mathrm{t}$ made them difficult to interpret and adopted an approximate frequentist inference to circumvent this issue.~\cite{McCarthy:2022gok} also noted a prior dependence of constraints on several cosmological parameters, especially depending on whether a uniform priors is taken on $\zeta$ or $\log_{10} \zeta$. In this section we show the effect of different priors on the marginalised posterior distributions using the profile likelihoods.

A uniform prior in $\log_{10} \zeta$, for example, corresponds to a prior on $\zeta$ proportional to $\zeta^{-1}$, and as we will see, leads to constraints that are entirely driven by the chosen lower bound on $\log_{10} \zeta$. In terms of an MCMC, the interpretation is that the sampler spends more time in the region of small $\zeta$ when the logarithm of the parameter is being sampled. To test the effect of this choice on the results of a Bayesian analysis, we have run MCMCs on the baseline data with uniform priors in four different parametrisations:

\begin{enumerate}[label=(\roman*)]
	\item $\zeta, \log_{10} a_\mathrm{t}, \kappa$, the standard in the rest of the paper,
	\item $\log_{10} \zeta, \log_{10} a_\mathrm{t}, \kappa$,
	\item $\zeta, \log_{10} a_\mathrm{t}, \log_{10} \kappa$, and finally
	\item $\log_{10} \zeta, \log_{10} a_\mathrm{t}, \log_{10} \kappa$,
\end{enumerate}
where a logarithmic lower bound $-3$ is taken for the parameters $\log_{10} \zeta $ and $\log_{10} \kappa$ which have physical boundaries at $0$.

The full posteriors obtained from these MCMCs can be seen in appendix~\ref{sec:apB}. It is seen that the effect on the $\Lambda$CDM parameters is relatively small, with the biggest shifts being in $h$ and $\omega_\mathrm{CDM}\equiv\Omega_\mathrm{CDM}h^2$ but still less than a standard deviation across all combinations of parametrisations. The constraints on the DMDR parameters, on the other hand, depend strongly on the chosen parametrisation.

\begin{figure*}
	\centering
	\includegraphics[width=0.8\textwidth]{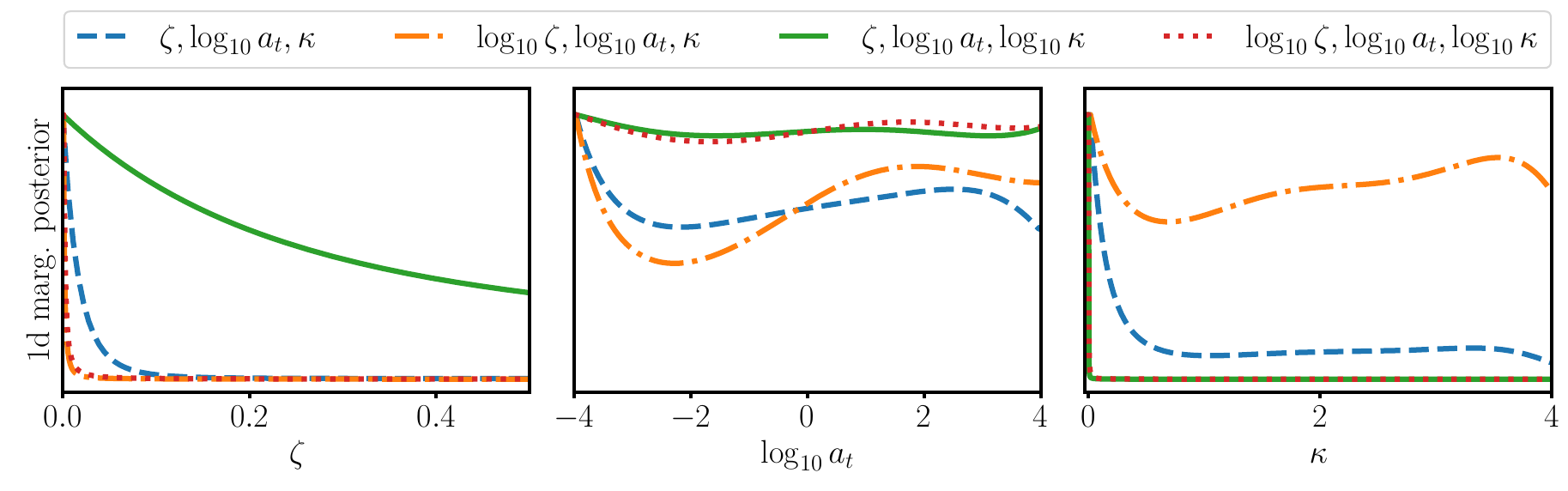}
	\caption{\label{fig:parametrisations} One-dimensional marginalised posterior distributions for the three DMDR parameters $\zeta$, $\log_{10} a_\mathrm{t}$ and $\kappa$, from runs with uniform priors in different parametrisations. Evidently, credible intervals obtained from these depend strongly on the chosen parametrisation. Profile likelihoods, on the other hand, are invariant under reparametrisations.}
\end{figure*}

\begin{table}
	\centering
	\begin{tabular}{?Sl|Sc|Sc|Sc?}
		\specialrule{.12em}{0em}{0em}
		\textbf{parametrisation} & $\zeta$ & $\log_{10} a_\mathrm{t}$ & $\kappa$ \\
		\hline
		$\zeta, \log_{10} a_\mathrm{t}, \kappa$ & $< 0.0334$ & unconstrained & $< 2.4000$ \\
		\hline
		$\log_{10} \zeta, \log_{10} a_\mathrm{t}, \kappa$ & $<0.0024$ & unconstrained & unconstrained \\
		\hline
		$\zeta, \log_{10} a_\mathrm{t}, \log_{10} \kappa$ & $< 1.1323$ & unconstrained & $< 0.0018$ \\
		\hline
		$\log_{10} \zeta, \log_{10} a_\mathrm{t}, \log_{10} \kappa$ & $< 0.0027$ & unconstrained & $< 0.0025$ \\
		\specialrule{.12em}{0em}{0em}
	\end{tabular}
	\caption{Bayesian $68 \%$ credible intervals for the three DMDR parameters $\zeta$, $\log_{10} a_\mathrm{t}$ and $\kappa$ subject to the baseline dataset but with uniform priors in a set of four different parametrisations of the model. }
	\label{tab:param}
\end{table}

Indeed, figure~\ref{fig:parametrisations} shows the one-dimensional marginalised posterior distributions on the parameters $\zeta$, $\log_{10} a_\mathrm{t}$ and $\kappa$ resulting from the four different parametrisations\footnote{For the runs that sampled a transformed version of these parameters, e.g.\ $\log_{10} \zeta$ instead of $\zeta$, we have transformed the posterior using the density transformation law.}, and table~\ref{tab:param} shows the $68.27 \ \%$ credible intervals constructed on the three parameters for each of the four different parametrisations. As an example, the upper bound on $\zeta$ shifts from $0.0334$ when using a uniform prior in $\zeta$ to $0.0024$ when using a uniform prior in $\log_{10}\zeta$. Thus, the final constraints obtained on these parameters depend on the chosen parametrisation. Since this is a phenomenological model, there exists no fundamental parametrisation, so there is no obvious choice of what parametrisation to choose. 

Although such a dependence on parametrisation is valid and may be desirable in some instances, it is difficult to \textit{a priori} know the extent to which a given credible interval is dominated by the parametrisation. As described in section~\ref{sec:2}, profile likelihoods are invariant under reparametrisation, so the frequentist analogue of table~\ref{tab:param} would have identical confidence intervals across the different parametrisations. Thus, the profile likelihood is a valuable tool to study the effects of the choice of parametrisations in a given model.

\subsection{Frequentist assessment of the Hubble tension}\label{sec:h}
\cite{McCarthy:2022gok} concluded, based on a Bayesian analysis, that the DMDR model does not alleviate the Hubble tension. However, several of the recently proposed solutions to the Hubble tension have recently been shown to provide a better resolution of the Hubble tension when analysed in the frequentist paradigm (e.g.~\cite{Holm:2022kkd} for the similar decaying cold dark matter model,~\cite{Herold:2021ksg, Herold:2022iib} for standard early dark energy and~\cite{Cruz:2023cxy} for new early dark energy). This is due to the fact that the proposed solutions are usually extensions of the $\Lambda$CDM model that have a control parameter, for example the abundance of a new species or the interaction strength of a novel interaction, upon the vanishing of which the $\Lambda$CDM cosmology is recovered. In this limit, then, the additional parameters of the extension become unconstrained, increasing the posterior volume around the vanishing of the control parameter, which acts to reduce the significance of the extension relative to the $\Lambda$CDM limit. Motivated by these examples, in this section we evaluate the ability of the DMDR model to alleviate the Hubble tension using profile likelihoods\footnote{Note the subtle point that although the global DMDR best-fit is in the $\Lambda$CDM limit and the profile likelihoods of the DMDR parameters are flat, the profile likelihood in $H_0$ may still be different from $\Lambda$CDM since for example the $\Delta N_\mathrm{eff}$ limit of the DMDR model is known to prefer slightly increased values of $H_0$~\citep{McCarthy:2022gok}.}. 
\begin{figure*}
	\centering
	\includegraphics[width=0.8\textwidth]{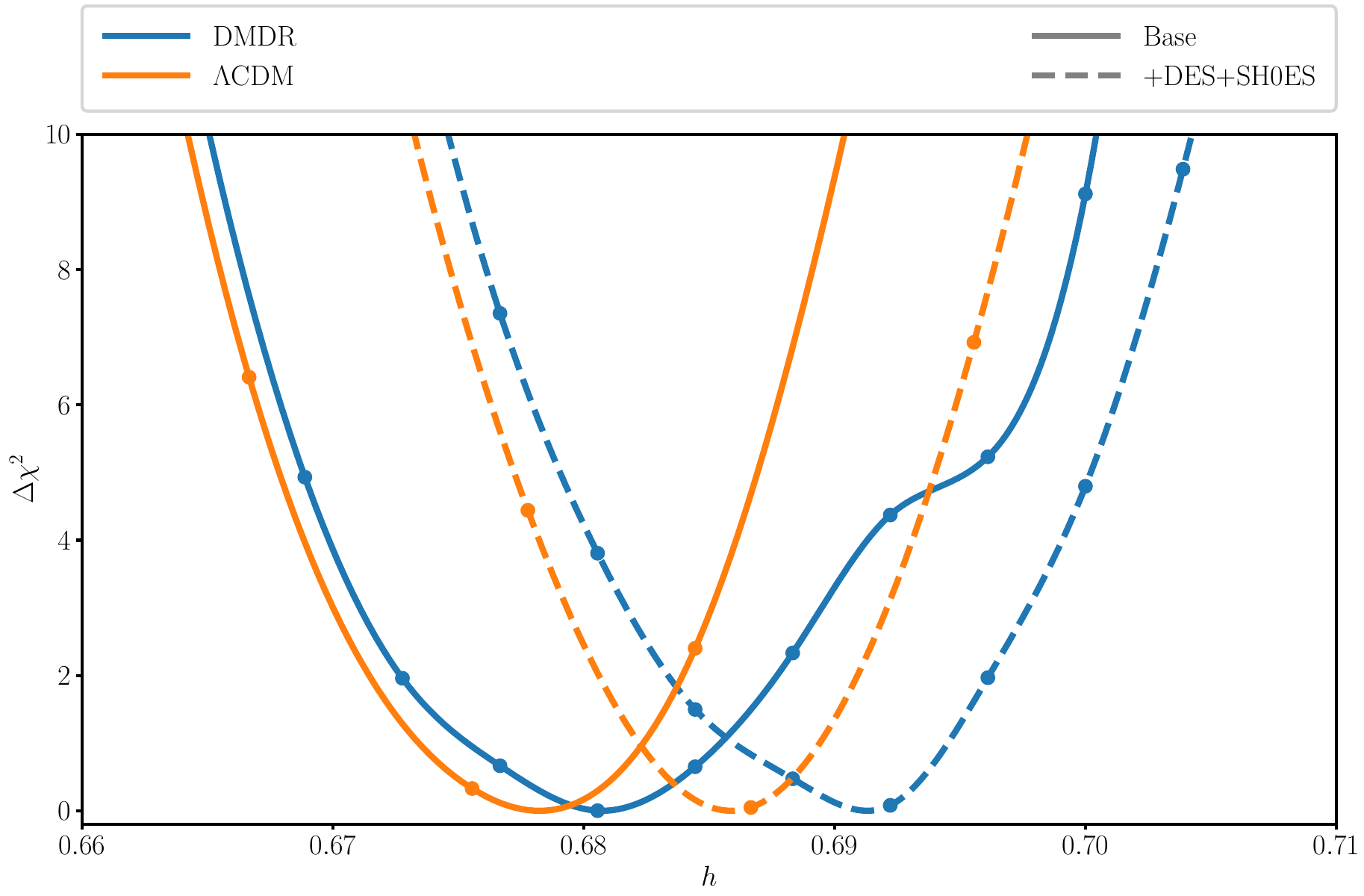}
	\caption{\label{fig:h_profiles} Profile likelihoods in terms of the quantity $\Delta \chi^2 \equiv -2 \log (L(h) / L_\mathrm{max})$ of the DMDR and $\Lambda$CDM models with the baseline dataset and additionally the DES and SH0ES priors.}
\end{figure*}

Figure~\ref{fig:h_profiles} shows the profile likelihoods, in terms of the $\Delta \chi^2$ quantity, of $h$ in the $\Lambda$CDM and DMDR models with the baseline dataset alone and including the DES and SH0ES described in section~\ref{sec:model}.

We see that the DMDR and $\Lambda$CDM predictions of $h$ with the baseline data is roughly the same, whereas a $\sim 1\sigma$ larger $h$ is obtained in the DMDR model when DES and SH0ES data is included. This is an indication of the naturally larger values of $H_0$ found in the DMDR model; however, we note that
this is still only a small alleviation, especially when compared to other models proposed to solve the $H_0$ tension~\citep{Schoneberg:2021qvd}. In conclusion, our findings, based on a frequentist inference of the DMDR model, corroborate the Bayesian result of~\cite{McCarthy:2022gok} that the model does not solve the Hubble tension\footnote{Although this may appear non-suprising, the difference between frequentist and Bayesian inference has lead to diverging conclusions in other potentially tension-solving models, such as early dark energy~\citep{Herold:2021ksg, Herold:2022iib}.}.

\begin{figure*}
	\centering
	\includegraphics[width=0.8\textwidth]{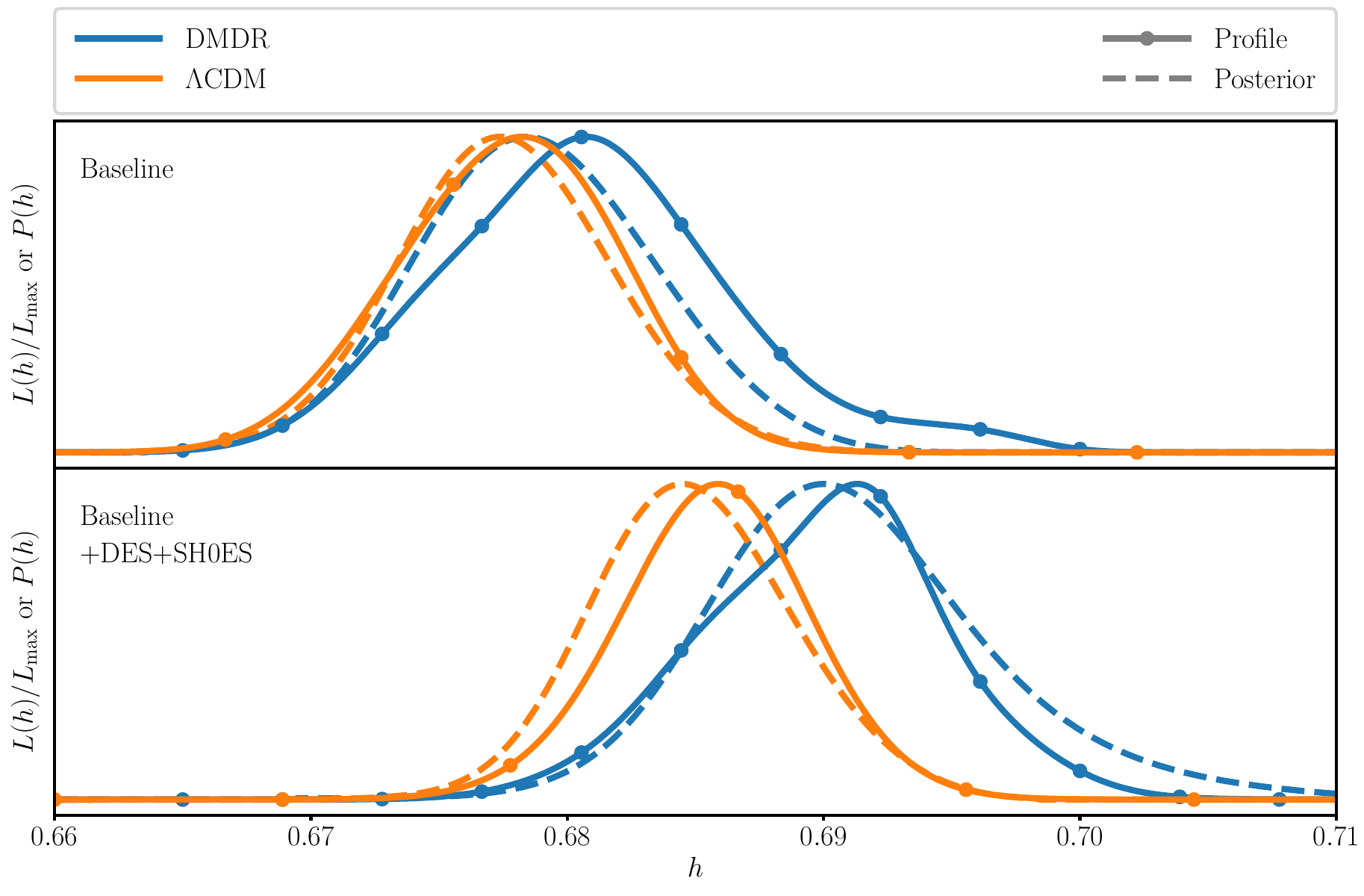}
	\caption{\label{fig:h} One-dimensional marginalised posteriors and profile likelihoods for the $\Lambda$CDM and DMDR models under baseline and baseline with priors on $\sigma_8$ and $h$ from the measurements of DES~\protect\citep{DES:2021wwk} and SH0ES~\protect\citep{Riess:2021jrx}. With the baseline data, there is a slight discrepancy between the DMDR posterior and profile likelihood due to a volume effect biasing the Bayesian DMDR inference toward the $\Lambda$CDM limit. When including a likelihood on the value of $h$ measured by the SH0ES collaboration, the volume effect disappears, and the posterior and profile likelihood agree. In either case, the model does not admit the large values of $h$ required to sufficiently alleviate the $H_0$ tension.}
\end{figure*}
Let us finally use the profile likelihoods to examine the extent to which the Bayesian constraints are dominated by volume effects. In figure~\ref{fig:h} we present posteriors and profile likelihoods on $h$ in the $\Lambda$CDM and DMDR models, respectively, using the baseline data and additionally the DES and SH0ES priors, respectively. The $\Lambda$CDM posterior and profile likelihoods match, as expected, since the $\Lambda$CDM posteriors in the baseline data are approximately Gaussian. With the baseline data, the DMDR posterior is almost identical to the $\Lambda$CDM posterior, whereas the profile likelihood prefers somewhat larger values of $h$. This is a manifestation of the strong volume effect in the $\Lambda$CDM limit which we discussed in section~\ref{sec:volume_effects}, which gives a bias of the posterior toward the $\Lambda$CDM values of the cosmological parameters. When SH0ES data is included, the MCMC sampler is drawn away from the $\Lambda$CDM limit of the DMDR model due to its preference for a larger $h$, largely removing the volume effect, which is seen, as the DMDR posterior and profile roughly match in this case. 

In conclusion, there is a slight volume effect in the baseline dataset where, in the Bayesian analysis, the DMDR model is drawn to its $\Lambda$CDM limit. We saw in section~\ref{sec:volume_effects} that this limit constitutes a strong volume effect that has large impacts on the DMDR parameters. Nevertheless, its impact on $h$ is mild, and not enough to change the conclusion that DMDR does not solve the Hubble tension.

\section{Conclusion}\label{sec:conclusion}
In this paper, we have introduced \prospect{}, a code for cosmological parameter inference using profile likelihoods. \prospect{} constructs an approximate profile likelihood from an MCMC, and optimises it using simulated annealing, a gradient-free stochastic optimisation algorithm. It employs an automatic tuning of the step size parameter and binned covariance matrices from the MCMC to achieve efficient optimisations of the profile likelihood. It interfaces with both \cobaya{} and \montepython{}, and is therefore fully compatible with modified versions of Einstein-Boltzmann solvers such as \class{} and \camb{} and the samplers \cobaya{} and \montepython{}. \prospect{} is publicly available at \url{https://github.com/AarhusCosmology/prospect\_public}.

Furthermore, we have illustrated the usefulness of profile likelihoods with a phenomenological model where dark matter is converted to dark radiation. This model reduces to the standard $\Lambda$CDM cosmology in numerous limits of its three additional model parameters. In these limits, the other parameters describing the conversion are unconstrained, leading to a large posterior probability mass, increasing the Bayesian favour for the $\Lambda$CDM limit. Although the profile likelihoods for these parameters are flat\footnote{Thus, to constrain the model parameters using profile likelihoods, two-dimensional profiles in pairs of the model parameters must be employed. Since these are more computationally demanding, we leave it for future work.}, since the global best-fit is in such a $\Lambda$CDM limit, we observe that this volume effect has a mild impact on the value of the Hubble parameter predicted by the model. We also highlight the impact of sampling linearly or logarithmically in the conversion parameters in the Bayesian analysis, and establish that constraints on these parameters are entirely driven by prior choices, although constraints on the $\Lambda$CDM parameters are only mildly affected. 

These examples illustrate how the Bayesian conclusions may be influenced by prior choices rather than by data. While this may be desired in some cases, it can be difficult to know how large the impact of prior choices are. In such cases, profile likelihoods provide exact assessments of prior effects, since they are reparametrisation and prior independent. Thus, although the conclusions from Bayesian and frequentist inferences have different interpretations, they illuminate the relationship between model and data from different useful viewpoints. Hence, instead of using either the Bayesian or frequentist approach, we believe that a combination of both is desirable for a fully nuanced statistical analysis.

\section*{Acknowledgements}

We thank Fiona McCarthy, J. Colin Hill, Felix Kahlhoefer, Torsten Bringmann and Kai Schmidt-Hoberg for useful comments on the draft. E.B.H. thanks Hans Kristian Eriksen for helpful discussions. We acknowledge computing resources from the Centre for Scientific Computing Aarhus (CSCAA). A.N., E.B.H., and T.T. were supported by a research grant (29337) from VILLUM FONDEN. 

\section*{Data Availability}

The \prospect{} code is publicly available at \url{https://github.com/AarhusCosmology/prospect\_public}. The MCMC chains and covariance matrices of the runs conducted in section~\ref{sec:3} will be shared on request. 
	
\bibliographystyle{mnras}
\bibliography{paper} 

\appendix

\section{The PROSPECT code}\label{sec:prospect}
\prospect{} is a code for cosmological inference written in Python. It can be installed by running the command \texttt{pip install prospect-public} in a terminal, and the source code can be found on GitHub\footnote{\url{https://github.com/AarhusCosmology/prospect\_public}} where the user will also find documentations and tutorials on how to use it. Its purpose is to take an MCMC made in \montepython{} or \cobaya{} and output a profile likelihood in one or more of the parameters being varied in the MCMC. During development, we focussed on making \prospect{} extremely user-friendly while also being fully computationally efficient. In section~\ref{sec:2}, we described the numerical strategies that make \prospect{} an \textit{efficient} code. In this section, instead, we highlight some of the most important features that make \prospect{} convenient to work with in practice.

\textbf{Plug-and-play interfacing with \montepython{} and \cobaya{}.} \prospect{} automatically reads and loads the likelihoods used in the MCMC it is initialised from. In order to be most widely applicable, we have interfaced \prospect{} with the two most popular cosmological inference codes, \montepython{} and \cobaya{}\footnote{\prospect{} also uses \getdist{}~\citep{Lewis:2019xzd} for internal analyses.}, in a most straightforward manner. The wrappers around \montepython{} and \cobaya{} are referred to as \textit{kernels} in the source code, and the purpose of the kernels is to define the full likelihood as a function of the cosmological parameters. For \montepython{}, this is accomplished by reading a \texttt{.conf} file defining paths to the \montepython{} and \class{} directories and importing the codes from there. For \cobaya{}, it simply imports \cobaya{} as a Python module and initialises the likelihood function through a \cobaya{} \texttt{model} object. These wrappers are constructed in a type of singleton object that ensures that \montepython{} and \cobaya{} are only ever initialised once per parallel process. The fact that \prospect{} does not use its own likelihood function but rather imports those used in the MCMC gives complete certainty that the likelihoods used are the same, admitting full comparisons between the two.

\textbf{Dynamical task system.} One of the main \textit{raisons d'être} of \prospect{}, which sets it apart from \montepython{} and \cobaya{}, is its dynamical workflow. We have broken the process of computing a profile likelihood into a set of \texttt{tasks} that are narrow in scope. \texttt{tasks} are stored in a prioritised task queue and can be completed in parallel. Crucially, \texttt{tasks} can add new \texttt{tasks} to the queue upon finishing: When \prospect{} is run, a single \texttt{InitialiseProfileTask} reads the values of the fixed parameter at which the user wants to evaluate the profile and launches an \texttt{InitialiseOptimiseTask} for each of the optimisations required in the profile. The latter constructs the initial points and covariance matrices, as described in~\ref{sec:initialisation} and launches an \texttt{OptimiseTask} equipped with these. The \texttt{OptimiseTask} carry out a small number of steps of an optimisation algorithm, checks for convergence, and emits a new \texttt{OptimiseTask} to the task queue that resumes the optimisation. Currently, \prospect{} only supports a \texttt{SimulatedAnnealing} optimiser, but it is written in a modular way such that it is straightforward to implement alternative optimisation algorithms in the future. 

In addition to the \texttt{OptimiseTask}s, an \texttt{AnalyseProfileTask} will be automatically submitted to the task queue at time intervals given by the user. These tasks browse the finished \texttt{OptimiseTask}s, creates a profile likelihood from them, and exports it to the disk along with summary statistics of the optimisation algorithms and interval constructions, if the user desires. Consequently, the user never has to analyse \prospect{} runs directly -- it is done automatically, in full parallel with the optimisations.

The task system is advantageous because it allows easy and efficient parallelisation of a very heterogeneous set of jobs. Furthermore, it is \textit{dynamic}, in the sense that \prospect{} can react to its own results and decide on its behaviour from that. This is already advantageous, since for example if the optimisation of any of the profile likelihood points converges, the final \texttt{OptimiseTask} will simply no longer submit a new \texttt{OptimiseTask} to the queue, essentially freeing up its process to work on other jobs in the queue. In the future, we plan on utilising this to implement adaptive methods of deciding which values to evaluate the profile at; something that is fairly non-trivial if the profile is unknown beforehand. Finally, the task system is fully modular, and it is straightforward for the user to define their own tasks by inheriting from a \texttt{BaseTask} object and following the example of the currently implemented tasks.

\textbf{Interactive snapshots.} \prospect{} does not rely on writing its results to disk on the go. Instead, the processes communicate internally, and the complete state of a \prospect{} run (a \textit{snapshot}) is saved as a pickled object at regular time intervals specified by the user, as a checkpoint to guard against crashes. This is useful because it means that all the information of a \prospect{} run is saved in a memory-efficient way. \prospect{} then has a wide range of functionalities to work with snapshots:
\begin{itemize}
	\item \textbf{Resuming from a snapshot} by reading the finished and queued tasks of the snapshot and simply continuing from those. This is useful for splitting up cluster jobs into smaller jobs that are easier to submit, for example.
	
	\item \textbf{Interactive snapshot loading.} In the \texttt{prospect} Python package, we have defined functions that load a \prospect{} snapshot in Python and gives easy access to the contents of the snapshot, which can be inspected manually and interactively. This is useful for debugging and to examine what exactly \prospect{} is doing with each \texttt{task}.
	
	\item \textbf{Interactive submission of new tasks.} At any one point, the user can load a snapshot, submit new \texttt{tasks} to the task queue, and save the snapshot, giving additional work to a \prospect{} run. This is useful for example if the user realises that the chosen binning of the profile likelihood is unfavourable, in which case the user can simply add new \texttt{InitialiseOptimiseTasks} by hand.
	
	\item \textbf{Reannealing.} In section~\ref{sec:simanneal}, we mentioned the concept of \textit{reannealing}, specific to simulated annealing, in which the temperature and step size are suddenly increased, with the aim of escaping from local minima and improving the global optimisation. We have defined a \texttt{prospect-reanneal} that submits a full set of new \texttt{OptimiseTask}s that start from the current best values, but with, potentially, a completely new optimisation algorithm or schedule. This is useful, since if the user discovers that their chosen schedule was not succesful in optimising satisfactorily, one can simply \texttt{reanneal} an existing snapshot instead of starting a completely new profile likelihood with the new schedule.
\end{itemize}

Lastly, all aspects of \prospect{} are written in a modular and cohesive way to allow for straightforward modifications in the future. We encourage anyone interested to try out \prospect{} and implement any bright ideas, whatever they may be\footnote{As described on the GitHub page, we are always happy to troubleshoot any issues or to aid in any new implementation.}.

\section{Full posteriors of the DMDR model}\label{sec:apB}
\begin{figure*}
	\centering
	\includegraphics[width=0.8\textwidth]{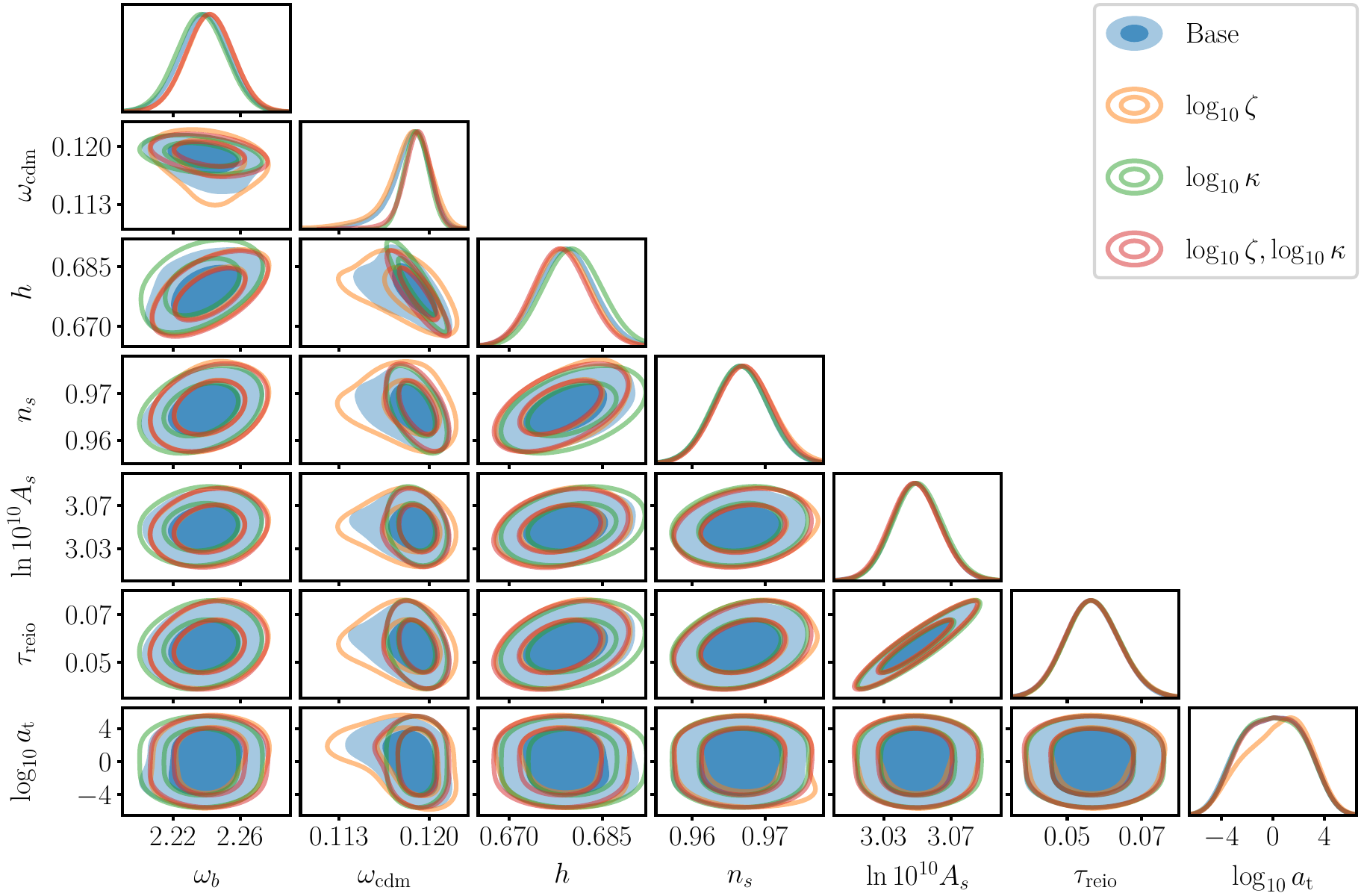}
	\caption{\label{fig:param_triangle} One and two dimensional marginalised posterior distributions of the DMDR model with the baseline dataset under the four different parametrisations presented in section~\ref{sec:prior_effects}.}
\end{figure*}

\begin{figure*}
	\centering
	\includegraphics[width=0.8\textwidth]{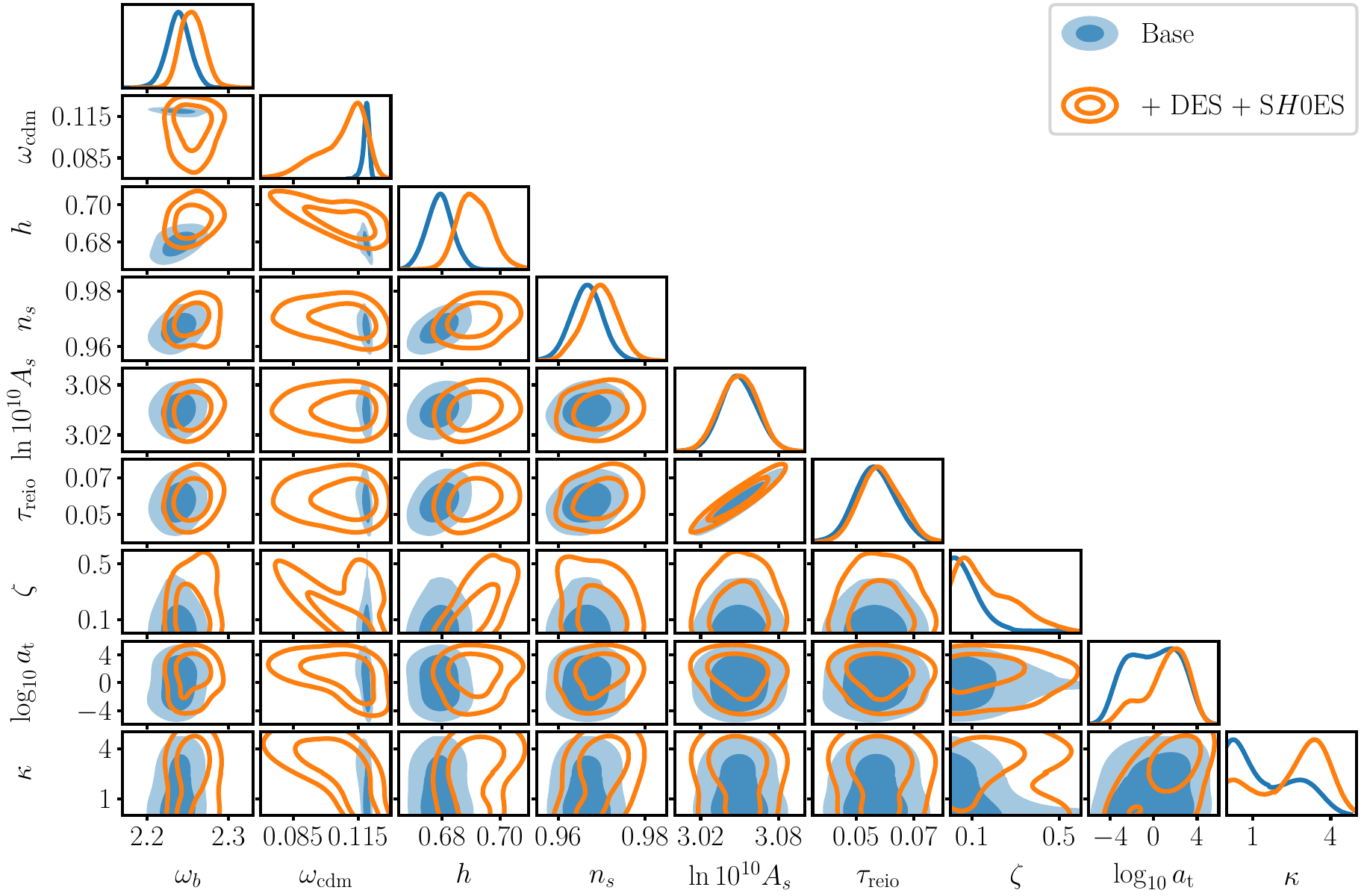}
	\caption{\label{fig:data_triangle} One and two dimensional marginalised posterior distributions of the DMDR model with the baseline dataset and additionally the DES and SH0ES priors.}
\end{figure*}

Figure~\ref{fig:param_triangle} shows a full triangle plot with the one and two dimensional marginalised posterior distributions over the 6 $\Lambda$CDM parameters for the four DMDR parametrisations discussed in section~\ref{sec:prior_effects} as well as the one DMDR parameter $\log_{10} a_t$ that the different parametrisations have in common.
Similarly, figure~\ref{fig:data_triangle} shows a full triangle plot of the parameters of the DMDR model subject to the baseline data set and additionally, the DES and SH0ES priors, as discussed in section~\ref{sec:h}.

\bsp	
\label{lastpage}
\end{document}